\documentclass{nic-series}

\begin{document}

\title{Phase Transitions and Relaxation Processes in Macromolecular Systems:
The Case of Bottle-brush Polymers}

\author{Hsiao-Ping Hsu \inst{1} 
\and Wolfgang Paul \inst{1,2}
\and Panagiotis E.~Theodorakis \inst{1}
\and Kurt Binder \inst{1}}

\institute{Institute for Physics,\\
           Johannes Gutenberg University Mainz, 55099 Mainz, Germany\\
         \email{\{hsu, wolfgang.paul, theodora, kurt.binder\}@uni-mainz.de}
         \and
           Institute for Physics,\\
           Martin-Luther University Halle-Wittenberg, 06120 Halle, Germany\\
         \email{wolfgang.paul@physik.uni-halle.de}
          }

\maketitle

\begin{abstracts}
As an example for the interplay of structure, dynamics, and phase
behavior of macromolecular systems, this article focuses on the
problem of bottle-brush polymers with either rigid or flexible
backbones. On a polymer with chain length $N_b$, side-chains with
chain length $N$ are endgrafted with grafting density $\sigma$.
Due to the multitude of characteristic length scales and the size
of these polymers (typically these cylindrical macromolecules
contain of the order of 10000 effective monomeric units)
understanding of the structure is a challenge for experiment. But
due to excessively large relaxation times (particularly under poor
solvent conditions) such macromolecules also are a challenge for
simulation studies. Simulation strategies to deal with this
challenge, both using Monte Carlo and Molecular Dynamics Methods,
will be briefly discussed, and typical results will be used to
illustrate the insight that can be gained.
\end{abstracts}

\begin{figure}[t]
\begin{center}
(a)\includegraphics[scale=0.60,angle=0]{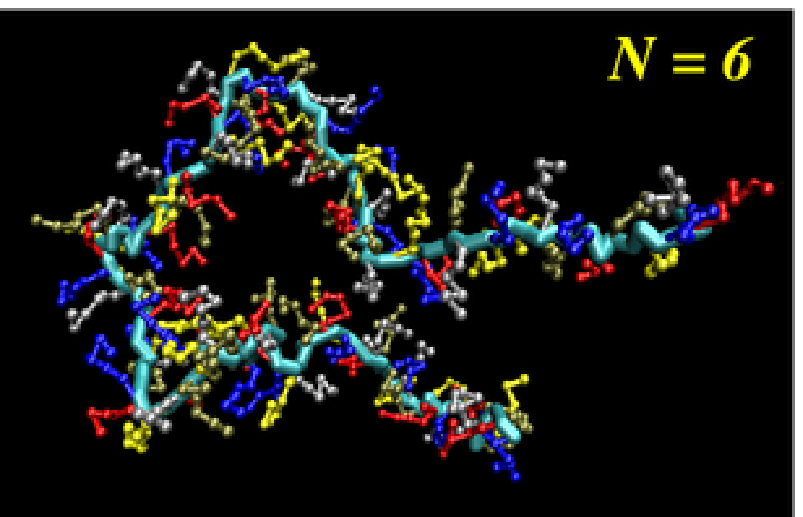}\hspace{0.4cm}
(b)\includegraphics[scale=0.60,angle=0]{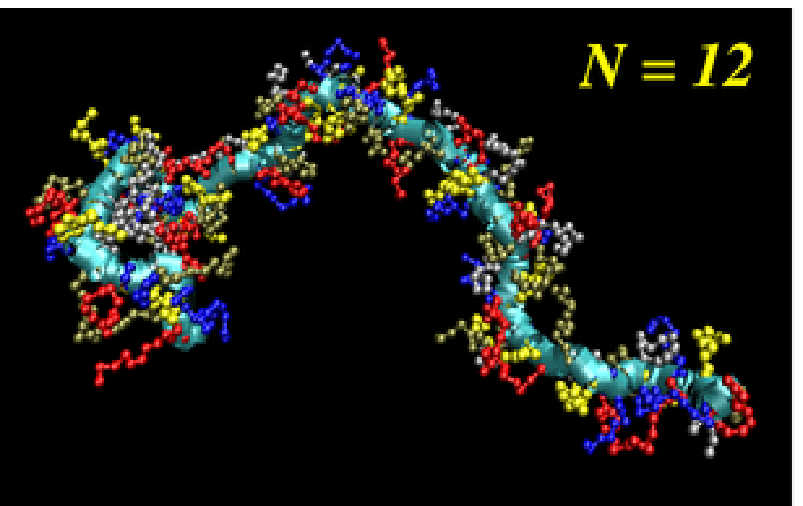}\\
\vspace{0.4cm}
(c)\includegraphics[scale=0.60,angle=0]{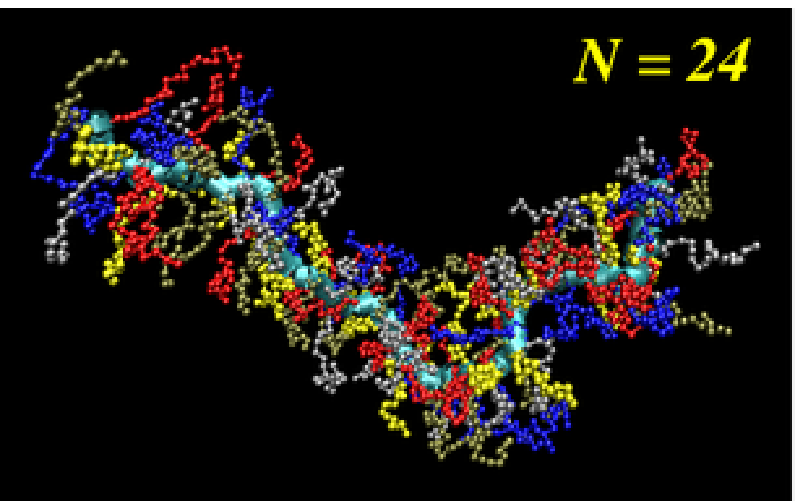}\hspace{0.4cm}
(d)\includegraphics[scale=0.60,angle=0]{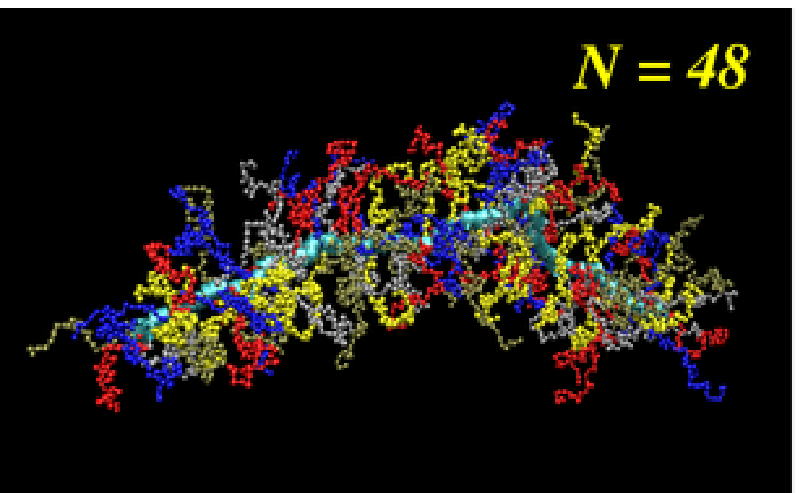}\\
\caption{\label{fig1} Snapshot pictures of simulated
bottle-brush polymers with a flexible backbone containing
$N_b=131$ effective monomers. At each backbone monomer one side
chain is grafted (grafting density $\sigma=1$) and side chain
lengths are $N=6$ (a), $N=12$ (b), $N=24$ (c) and $N=48$ (d). The
chains are described by the bond fluctuation model on the simple
cubic lattice (see Sec. 2) and the only interaction considered is
the excluded volume interaction between effective monomers.}
\end{center}
\end{figure}

\begin{figure}[t]
\begin{center}
\includegraphics[width=5.0cm,angle=0]{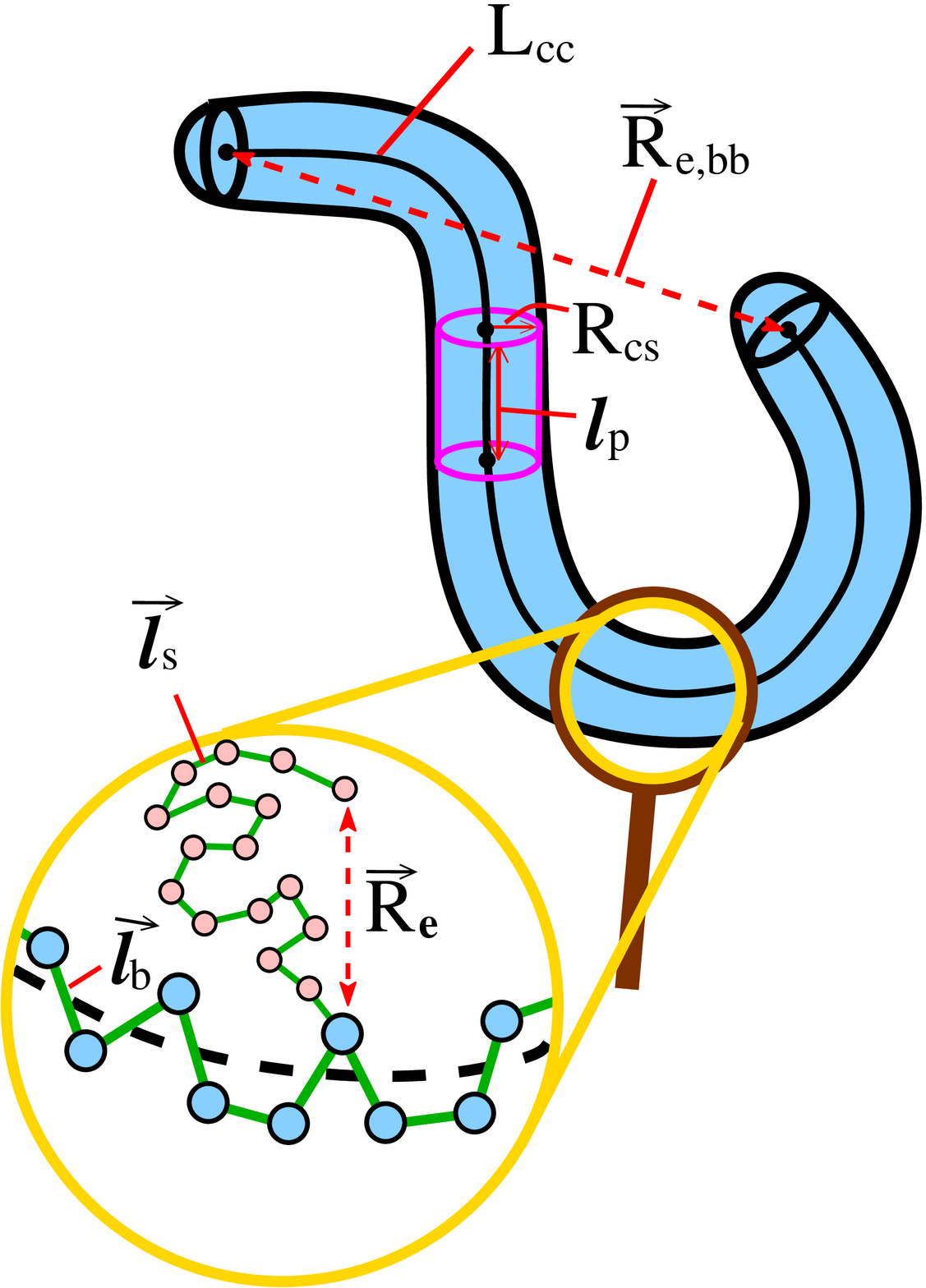}
\caption{\label{fig2} Multiple length scales are needed
for the structural characterization of a bottle-brush polymer: a
coarse-grained view describes this object as a flexible
spherocylinder of length $L_{cc}$ and cross-sectional radius
$R_{\rm cs}$, which is locally straight over a length $\ell_p$
(``persistence length''). A further length scale of interest is
the end-to-end distance $\vec{R}_{e,bb}$ of the backbone chain.
This coarse-grained view is experimentally obtained by 
atomic force microscopy imaging 
of bottle-brush polymer adsorbed on substrates,
for instance. A ``microscope'' with resolution on the scale of
atoms would see the monomers of the backbone
chains (connected by backbone bond vectors $\vec{\ell}_b$) and
monomers of the side chains (connected by bond vectors
$\vec{\ell}_s$). A mesoscopic length of interest then is the
end-to-end distance $\vec{R}_e$ of the side chains.}
\end{center}
\end{figure}

\section{Introduction}
The so-called ``bottle-brush polymers'' consist of a long
macromolecule serving as a ``backbone'' on which many flexible
side chains are densely grafted \cite{1,2}. Since the chemical
synthesis of such complex polymeric structures has become
possible, these molecular bottle-brushes have found much interest
for various possible applications: the structure reacts very
sensitively to changes of solvent quality (due to change of
temperature of the solution, pH value, etc.), enabling the use of
these molecules as sensors or actuators on the nanoscale
\cite{3,4}. For some conditions these bottle-brush molecules
behave like stiff cylindrical rods, and hence they can serve as
building blocks of supramolecular aggregates, or show orientational 
order in solution as nematic liquid crystals do. On the other
hand, these molecular bottle-brushes are also very ``soft'', i.e,
they show only very small resistance to shear deformation;
bottle-brush molecules of biological origin such as aggrecan which
occurs in the cartilage of mammalian (including human!) joints are
indeed held responsible for the excellent lubrication properties
(reducing frictional forces) in such joints \cite{5}.

However, for being able to control the function of these complex
macromolecules one must be able to control their structure, i.e.,
one must understand how the structure depends on various
parameters of the problem: chain length $N_b$ of the ``backbone'',
chain length $N$ of the side chains, grafting density $\sigma$ of
the side chains along the backbone, solvent quality, just to name
the most important ones of these parameters. This is the reason
why computer simulations are needed in order to understand such
bottle-brushes: even in the case of good solvent, the structure of
these objects is very complicated as the snapshot pictures
\cite{6,7} (Fig.~\ref{fig1}) of the simulated models show, and
there occur a multitude of characteristic length scales
(Fig.~\ref{fig2}) \cite{6,7} that are needed to describe the
structure; it is very difficult to extract this information from
experiment, and hence simulations are valuable here as they can
yield a far more detailed picture.

Of course, the large size of these bottle-brush polymers (even the
coarse-grained models described in the following sections contain
of the order of several thousand or even more than 10000 effective
monomers per polymer) is a serious obstacle for simulation, too;
in fact, developing efficient models and methods for the
simulation of macromolecular systems is a longstanding and
important problem \cite{8,9}. We hence shall address this issue of
proper choice of both suitable models and efficient algorithms in
the next sections.

\begin{figure}[t]
\begin{center}
(a)\includegraphics[scale=0.20,angle=270]{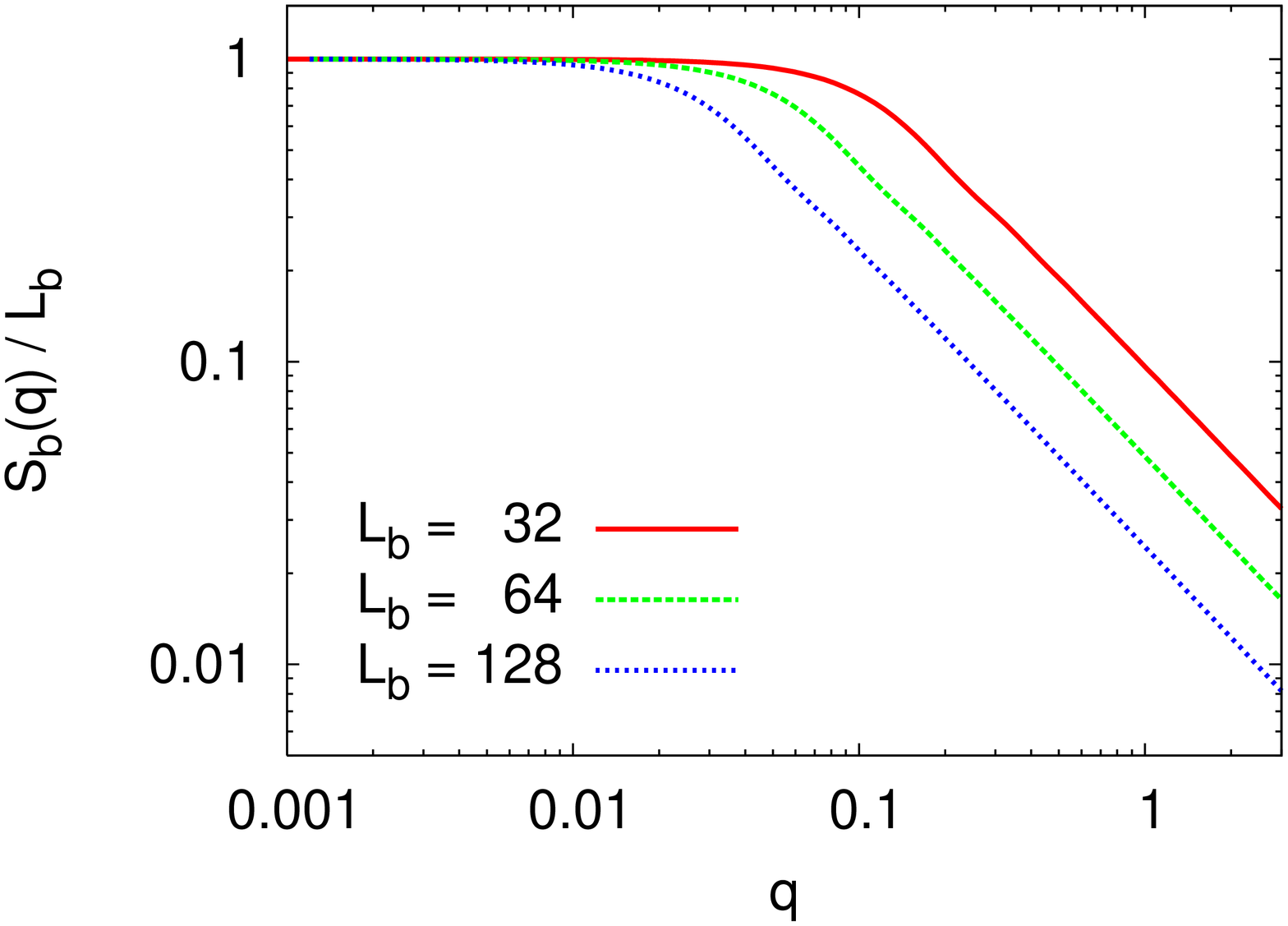}\hspace{0.4cm}
(b)\includegraphics[scale=0.20,angle=270]{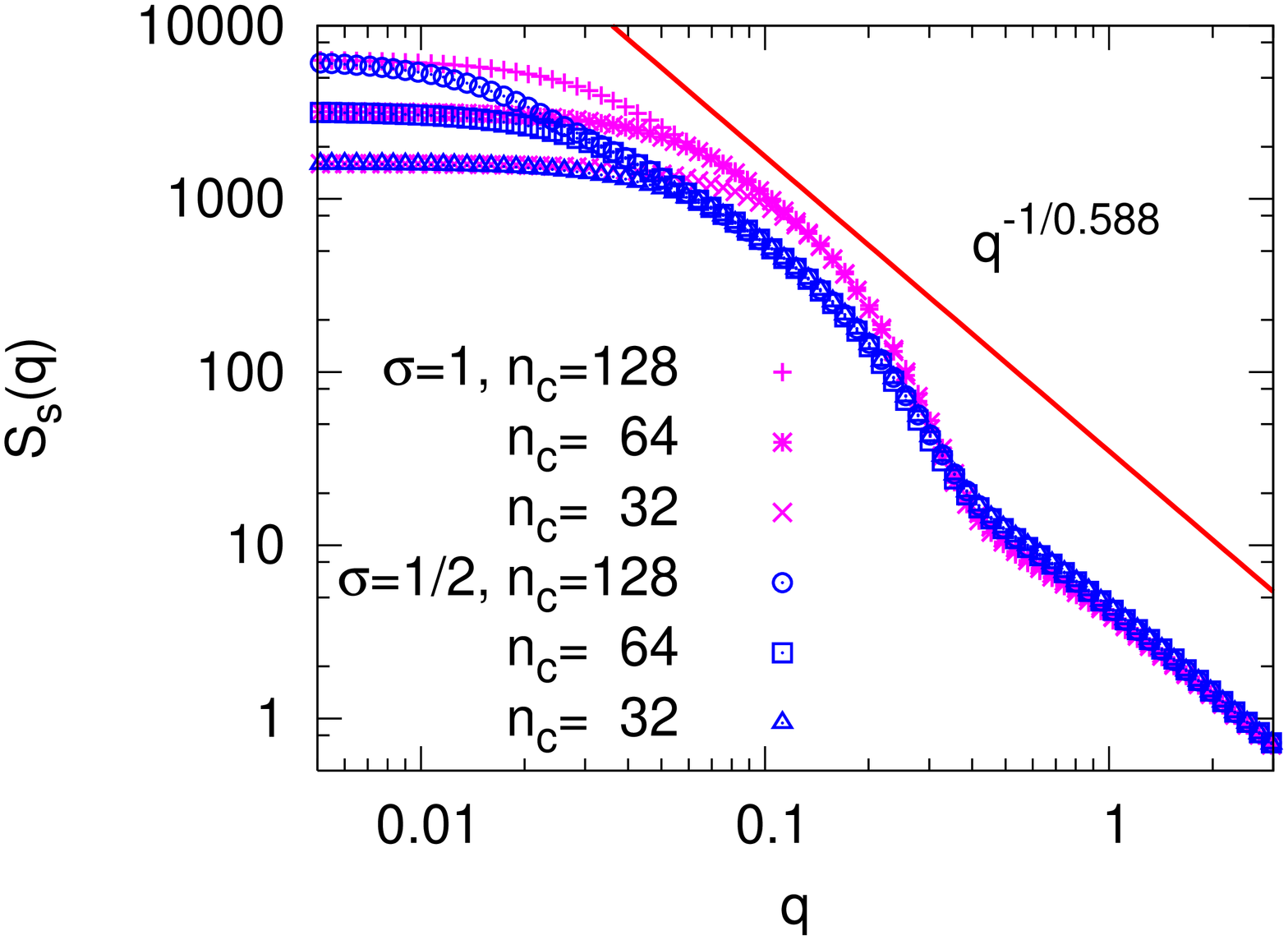}\\
\vspace{0.4cm}
(c)\includegraphics[scale=0.20,angle=270]{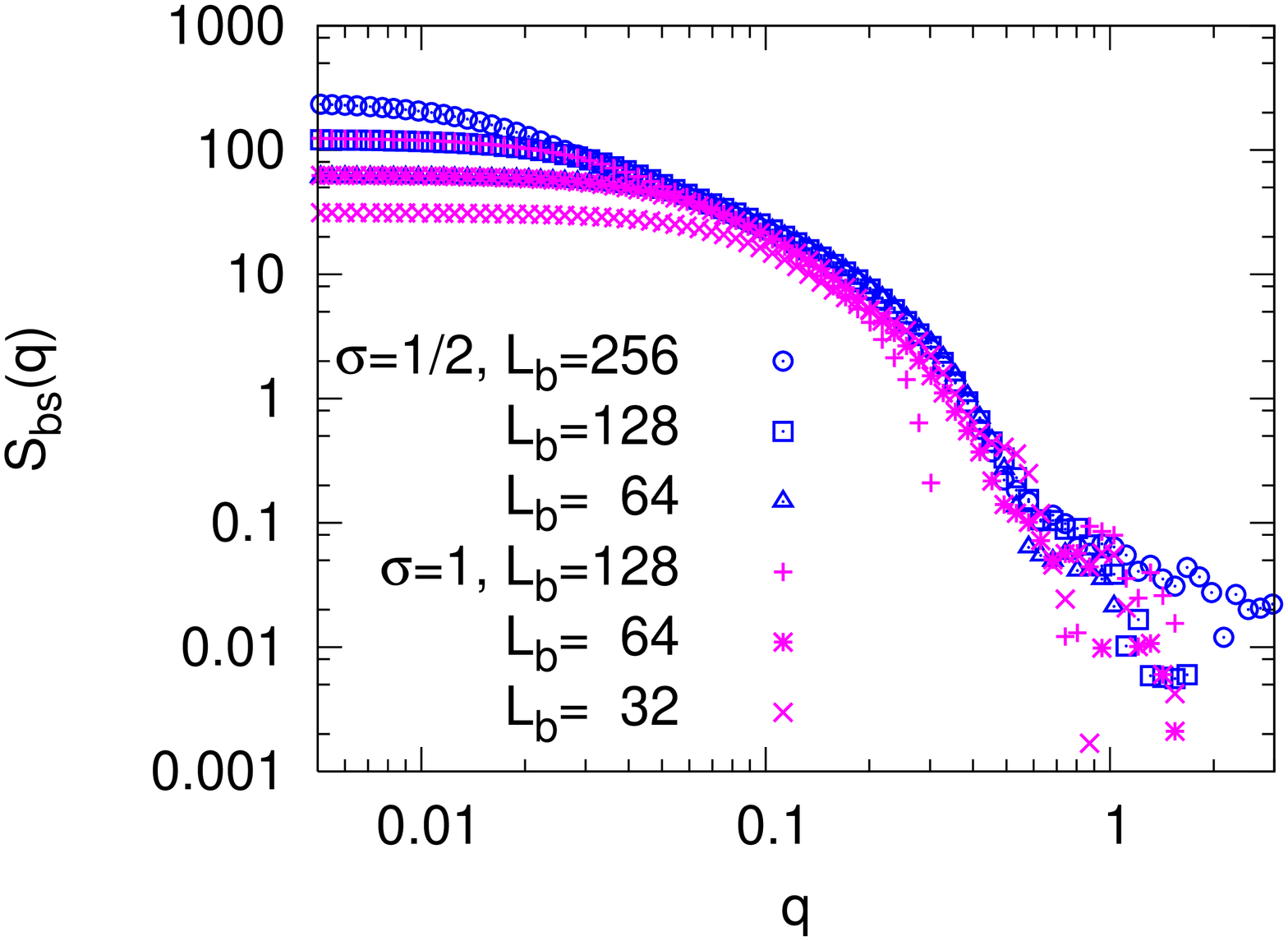}\hspace{0.4cm}
(d)\includegraphics[scale=0.20,angle=270]{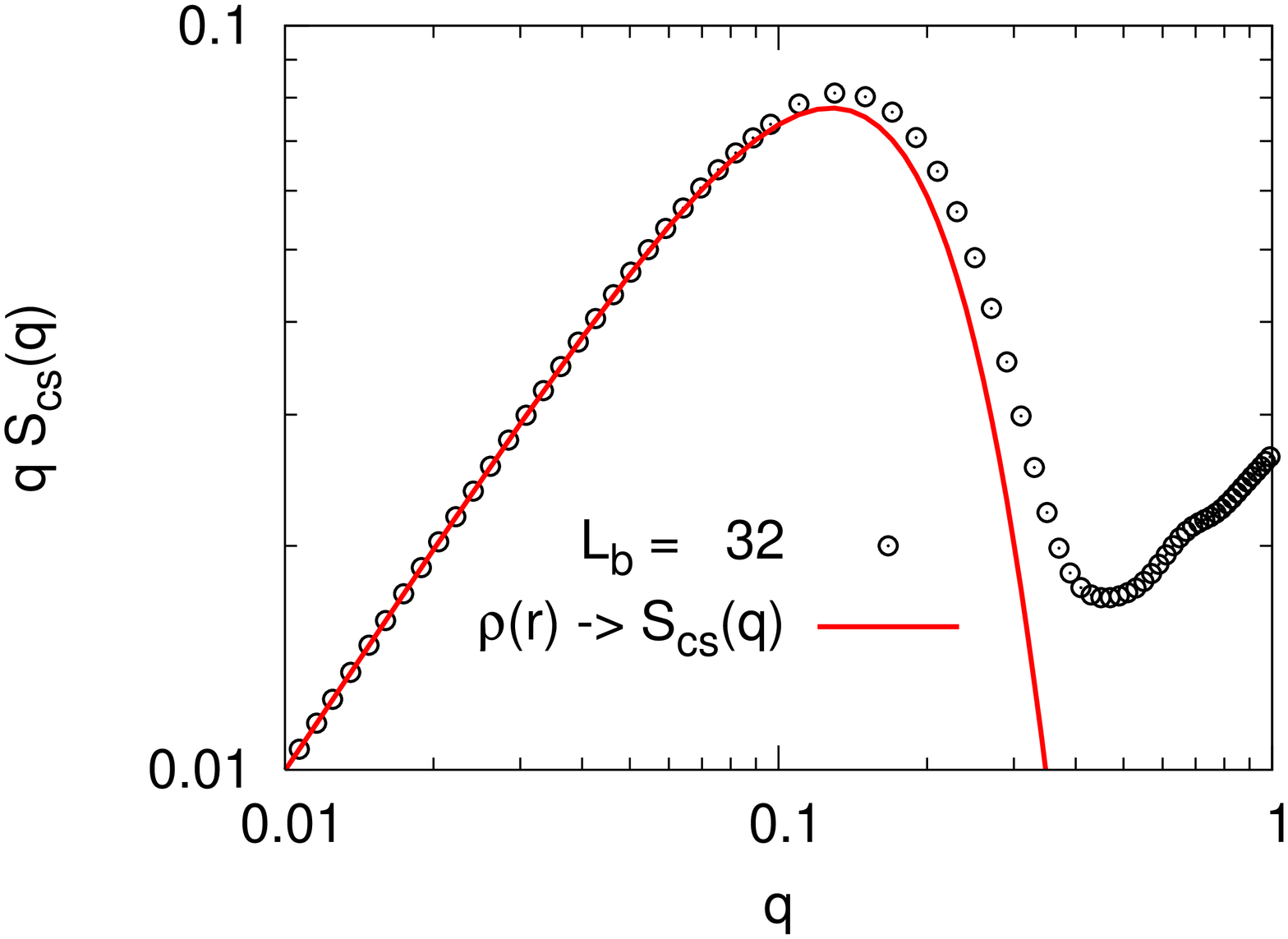}\\
\caption{\label{fig3} a) Log-log plot of the normalized
scattering function of the backbone $S_b(q)/L_b$, $L_b$ being the
backbone length, $L_b=N_b a$ where $a(=1)$ is the lattice spacing.
Three choices of $L_b$ are shown. b) Log-log plot of the scattering
from all the side chain monomers, for $N=50$, and two choices of
the grafting density c) Log-log plot of the scattering due to
interference contributions from monomers in side chains and the
backbone (for the same parameters as in b). d) Log-log plot of
$qS_{\rm cs}(q)$ vs. $q$ in the range $0.01 \leq q \leq 1$, comparing the
actual scattering due to the side chains with the prediction
resulting from Eq.~(\ref{eq2}), using $\rho(r)$ as actually
recorded in the simulation~\cite{12}.}
\end{center}
\end{figure}

\section{Bottle-brush polymers with rigid backbones in good
solvents}

If one assumes the backbone of the bottle-brush to be completely
stiff, the problem is reduced to grafting side chains of length
$N$ to a straight line. Although this limiting case seems somewhat
artificial, from the point of view that one wishes to model real
systems, it is a very useful test case: there is no reason to
assume that approximations that already fail in this rather simple
limit become accurate for the more complicated case of flexible
backbones; moreover, this case is rather simple to simulate, and
the analysis of the simulation data is relatively straightforward.

For the study of this case, polymers were simply represented as
standard self-avoiding walks on the simple-cubic lattice (i.e.,
beads of the chains are occupied lattice sites, connected by
nearest-neighbor links on the lattice, and multiple occupancy of
sites is forbidden \cite{8,9}).

As a simulation method, pruned-enriched Rosenbluth methods (PERM)
\cite{10,11} were used \cite{12}. In the Rosenbluth method, all
side chains are grown simultaneously step by step, choosing only
from sites which are not yet occupied, and the statistical weight
of the polymer configuration is computed recursively. In the PERM
algorithm, one does not grow a single polymer at a time, but a
large ``population'' of equivalent chains is grown simultaneously
but from time to time configurations with very low statistical
weight are killed, and configurations with large statistical
weight are ``cloned'' (``go with the winners''strategy \cite{10}).
The advantage is that (unlike dynamic Monte Carlo algorithms
\cite{8,9}) this method does not suffer from ``critical slowing
down'' when $N$ gets large, and data for all $N$ (up to the
maximum value studied) are obtained in a single simulation (see
\cite{11} for more details).

Now one hypothesis popular in most experimental studies (e.g.
\cite{13,14}) is the factorization approximation \cite{15} for the
structure factor $S(q)$ (that describes the small angle scattering
intensity of neutrons, X-rays or light from dilute solutions of
bottle-brushes polymers) into a contribution due to the backbone
($S(q)$) and due to the side chains $(S_{s}(q))$

\begin{equation} \label{eq1}
S(q) \approx S_b(q) S_s (q) \approx S_b(q) S_{\rm cs}(q)
\end{equation}
where $q$ is the absolute value of the scattering wave vector. 
In the last step the side chain scattering was 
approximated by $S_{\rm cs}(q)$. This cross sectional scattering
is in turn approximated related to the radial density $\rho_{\rm cs}(r)$
perpendicular to the cylinder axis at which the side chains are
grafted as (the constant $c$ ensures proper normalization
\cite{12})

\begin{equation} \label{eq2}
S_{\rm cs}(q) = c^{-1} \langle |\int d^2 \vec{r} \rho_{\rm cs} (r) \exp (i
\vec{q} \cdot \vec{r})|^2 \rangle .
\end{equation}
Obviously, Eq.~(\ref{eq1}) neglects interference effects due to
correlations between the monomer positions in the side chains and
in the backbone. Writing the scattering due to the monomers in the
side chains as the average of a square \{Eq.~(\ref{eq2})\} ignores
correlations in the occupation probability in the $z$-direction
along the bottle-brush axis. All such correlations do contribute
to the actual structure factor, of course, when it is computed
from its definition,

\begin{equation} \label{eq3}
S(q) =\frac{1}{\mathcal{N}_{\rm tot}} \,
\sum\limits_{i=1}^{\mathcal{N}_{\rm tot}} \,
\sum\limits_{j=1}^{\mathcal{N}_{\rm tot}} \langle c(\vec{r}_i)
c(\vec{r}_j) \rangle \sin
(q |\vec{r}_i-\vec{r}_j|)/(q|\vec{r}_i-\vec{r}_j|) \quad .
\end{equation}
Here $c(\vec{r}_i)$ is an occupation variable, $c(\vec{r}_i)=1$ if
the site $i$ is occupied by a bead, and zero otherwise, and the
sums extend over all $\mathcal{N}_{\rm tot}$ monomers (in the side
chains and in the backbone).

In the simulation, one can go beyond experiment by not only
recording the total scattering $S(q)$, but also the contributions
$S_b(q)$, $S_s(q)$ and $S_{\rm cs}(q)$ individually (Fig.~\ref{fig3}).
One can see that the approximation $S(q) \approx S_b(q) S_s(q)$
leads to a relative error of the order of a few \% only (cf. the
difference in ordinate scales between parts b) and c), while the
assumption $S_s(q)\approx S_{\rm cs}(q)$ [with Eq.~(\ref{eq2})] leads
to appreciable errors at large $q$ (see part d)). Therefore the
use of Eqs.~(\ref{eq1}),~(\ref{eq2}) to analyze experiments can
lead to appreciable errors, when one wishes to predict \cite{12}
the radial density distribution $\rho(r)$.

\begin{figure}[t]
\begin{center}
\includegraphics[width=5cm,angle=270]{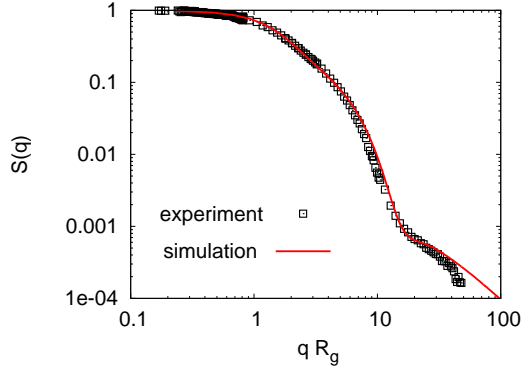}
\caption{\label{fig4} Structure factor $S(q)$ of
bottle-brush polymer obtained \cite{13} from scattering experiment
mapped to the simulated model (cf. text) by requiring that the
total gyration radius $R_g$ is matched, to fix the translation
factor for the length scale \cite{7}.}
\end{center}
\end{figure}

\begin{figure}[t]
\begin{center}
\includegraphics[width=5.0cm,angle=270]{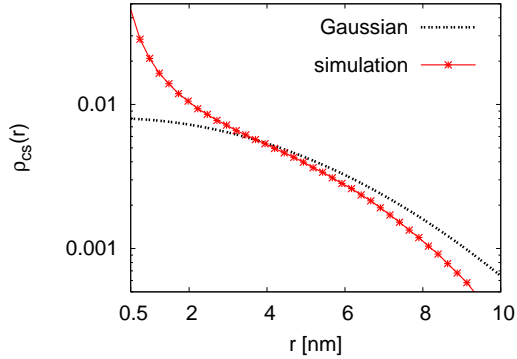}
\caption{\label{fig5} Log-log plot of the density
profile $\rho_{\rm cs}(r)$ vs.~$r$, for the systems shown in
Fig.~\ref{fig4}. For fitting the experimental data, a Gaussian
form for $\rho_{\rm cs}(r)$ was assumed \cite{13}.}
\end{center}
\end{figure}

\section{Bottle-brushes with flexible backbones in good solvents}

In the case of flexible backbones, the PERM algorithm no longer
provides an efficient sampling of configuration space; there
``dynamic'' Monte Carlo algorithms are used, but with ``unphysical
moves'' allowing chain intersections and large configurational
changes in a single step, as achieved by the Pivot algorithm
\cite{8,9}. Using the bond fluctuation model \cite{8,9}, the ``L26
algorithm'' \cite{16} was used \cite{17} for local moves allowing
bond intersection and nevertheless respecting excluded volume. In
this way well-equilibrated data for systems with up to $N_b=259$
monomers at the backbone and up to $N=48$ monomers in the side
chains could be studied (for $\sigma=1$). Examples of snapshot
pictures have already been given in Fig.~\ref{fig1}.

Adjusting the physical meaning of the lattice spacing to
correspond to 0.263 nm, the structure factor of the simulated
model matches almost perfectly an experimental result \cite{13}
(for slightly different numbers of chemical monomers,
$N^{\exp}_b=400$, $N^{\exp}=62$; this difference is irrelevant,
since there is no one-to-one correspondence between covalent bonds
and the ``effective bonds'' of the model) see \cite{7}
Fig.~\ref{fig4}. From the simulation one can directly extract the
cross-sectional density profile $\rho_{\rm cs}(r)$ and compare it
\cite{7} to the approximate experimental result
(Fig.~\ref{fig5}), which was obtained from fitting $S(q)$ to
Eqs.~(\ref{eq1}),~(\ref{eq2}) \cite{13}. One sees that the
analysis of the experiment  could predict roughly correctly the
distance on which the profile $\rho_{\rm cs}(r)$ decays to zero, but
does not account for its precise functional form. A further
interesting finding \cite{7,17} is the result that the
persistence length $\ell_p$ (cf. Fig\ref{fig2}) depends strongly
\cite{7} on $N_b$ (at least if one uses the textbook definitions
\cite{19}), and hence is not a useful measure of local chain
stiffness.

\begin{figure}[t]
\begin{center}
\includegraphics[width=7.0cm,angle=0]{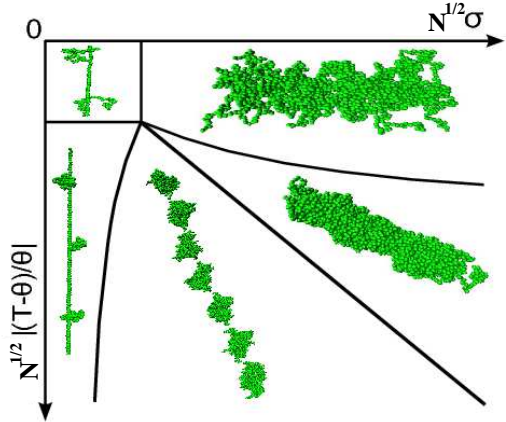}
\caption{\label{fig6} Schematic diagram of states of a
bottle-brush polymer with a rigid backbone under poor solvent
conditions in the plane of variables $x=N^{1/2} \sigma$ and
$y=N^{1/2}(1-T/\theta)$, as proposed by Scheiko et al. \cite{20}.
The lines indicate (smooth) crossovers between the different
states \cite{20}, which we have characterized by snapshots from
our simulations \cite{21}. For further explanations see text.}
\end{center}
\end{figure}

\section{Bottle-brushes in poor solvents}

While isolated single chains in dilute solution collapse to dense
globules \cite{19} when the temperature $T$ is lowered below the
Theta temperature $T=\theta$, for a bottle-brush the constraint of
the chains being grafted to a backbone leads to an interesting
diagram of states, considering $1-T/\theta$ and $\sigma$ as
variables (Fig.~\ref{fig6}) \cite{20}. Near $T=\theta$ and $\sigma
N^{1/2} \rightarrow 0$ the side chains collapse to dense globules 
when $N^{1/2}(1-T/\theta) \gg 1$.
For $T \approx \theta$ and $N^{1/2} \sigma \gg 1$ one has a
cylindrical structure (as in the previous sections) which
collapses to a dense cylinder when $N^{1/2} (1-T/\theta) \gg 1$.
However, interestingly at intermediate grafting densities and
$N^{1/2} (1-T/\theta) >1$ a laterally inhomogeneous ``pearl
necklace-structure'' was predicted \cite{20}.

\begin{figure}[t]
\begin{center}
\includegraphics[width=7.0cm,angle=270]{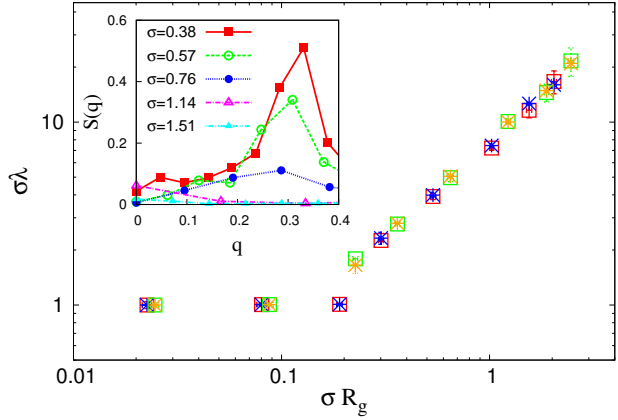}
\caption{\label{fig7} Log-log plot of wavelength
$\lambda$ (dimensionless by normalization with $\sigma$) versus
grafting density $\sigma$ (dimensionless by normalization with the
radius of gyration of the side chains, $R_g$).
The wavelength was extracted either from the
$z$-dependence of a normalized correlation function at radial
distance $r\approx3$ Lennard-Jones units $\sigma$ (stars) or
extracted from the peak position of its Fourier transform $S(q)$
(squares) [inset]. All data refer to temperature $T=1.5$
(in Lennard Jones units). Blue and red symbols refer to $N=35$, 
orange and green ones
to $N=50$. Inset shows $S(q)$ vs. $q$ for $N=35$ and grafting
densities $\sigma \geq 0.38$.}
\end{center}
\end{figure}

This problem can neither be studied efficiently by the PERM method
nor by the bond fluctuation algorithm of Sec. 3 - both methods
get very inefficient for dense polymer configurations. Thus,
instead Molecular Dynamics simulations of the standard
Grest-Kremer-type bead-spring model \cite{8,9} were carried out
\cite{21}. Fig.~\ref{fig7} shows \cite{21} characteristic
wavelength $\lambda$ plotted vs. grafting density $\sigma$. One
finds that for small $\sigma$ a trivial periodicity given by
$1/\sigma$ occurs, while for $\sigma R_g \geq 0.2$ 
($R_g = \langle R^2_g \rangle^{1/2}$)
the periodicity is essentially independent of $\sigma$,
given by about that $\lambda \approx 8 R_g$. Of course, mapping
out the phase behavior precisely for a wide range of $N$, $\sigma$
and $T$ is a formidable problem, due to very long relaxation times
of the collapsed states in Fig.~\ref{fig6}.

\section{Concluding Remarks}
The above examples have illustrated that polymers with complex
architecture pose challenging problems with respect to their
structure and thermodynamics. Investigation of their relaxation
behavior is another interesting problem, to be tackled in the
future. The above examples have also illustrated the need to
carefully adjust both the model and the algorithm to the problems
one wants to deal with.

\section*{Acknowledgments}
One of us (H.-P. Hsu) received support from the DFG (SFB625/A3),
another (P.~E. Theodorakis) from the MPG via a Max-Planck Fellowship of the
MPI-P. We are grateful to the NIC J\"ulich for computer time at
the JUMP and SoftComp cluster.

\end{document}